\documentclass[a4paper]{spie}  
\usepackage[]{graphicx}

\title{Optical Turbulence Forecast: towards a new era of the ground-based astronomy} 

\author{Elena Masciadri\supit{a}\skiplinehalf
\supit{a}INAF - Osservatorio Astrofisico di Arcetri, L.go E. Fermi 5, 50125  Florence, Italy \\
}

\authorinfo{E-mail: masciadri@arcetri.astro.it, Telephone: +39 055 2752202\\  }
 
\def\LO{\mbox{${\cal L}_0 \ $}}
\def\CN2{\mbox{$C_N^2 \ $}}

\def\CT2{\mbox{$C_T^2 \ $}}

\def\sigmaI2{\mbox{$\sigma ^{2}_{I} \ $}}

\begin{document} 
  \maketitle 

\begin{abstract}
The simulation of the optical turbulence (OT) for astronomical applications 
obtained with non-hydrostatic atmospherical 
models at meso-scale presents, with respect to measurements, some advantages.
Among these:
{\bf (1)} the possibility to provide 3D $\CN2$ maps above a region of a few tens 
of kilometers around a telescope.  
{\bf (2)} the possibility to simulate the turbulence {\it 'where'} and {\it 'when'} it is 
desired without the need of long and expensive site testing campaigns 
done with several instruments.  
{\bf (3)} the possibility to forecast the optical turbulence, goal
considered a 'chimera' by all astronomers and fundamental element for 
the implementation of the flexible-scheduling, crucial operation mode for the  success of new class of telescopes (D $>$ $10$ $m$)

The future of the ground-based astronomy relies upon the potentialities and 
feasibility of the ELTs. Our ability in knowing,
controlling and 'managing' the effects of the turbulence on such a 
new generation 
telescopes and facilities are determinant to assure their competitiveness 
with respect to the space astronomy.
 
In the past several studies have been carried out proving
the feasibility of the simulation of realistic $\CN2$ profiles 
above astronomical sites. 
The European Community (FP6 Program) decided recently to fund a 
Project aiming, from one side, to 
prove the feasibility of the OT forecasts and the ability of meso-scale 
models in discriminating astronomical sites from optical turbulence point of view and, from the other side, 
to boost 
the development of this discipline at the borderline between the 
astrophysics and the meteorology.

In this contribution I will present the scientific and technological 
goals of this project, the challenges for the ground-based astronomy 
that are related to the success of such a project and the international 
synergies that have been joint to optimize the results.
\end{abstract}




\section{TURBULENCE FORECASTS AND CHALLENGES FOR GROUND-BASED ASTRONOMY}
\label{sect:challanges} 

The future of the ground-based astronomy relies upon the potentialities and feasibility of the new class of Extremely Large Telescopes (ELTs) (D $\ge$ $30$ m). In spite of the presence of the atmospherical turbulence, the ground-based observations are still quite competitive with respect to the spatial ones due to the lower financial investment, a longer typical lifetime of telescopes and a better angular resolution attained thanks to the larger pupil sizes of the ground-based telescopes. The latter can be obtained with the employment of the Adaptive Optics (AO) techniques, expressly conceived to reduce/eliminate perturbations produced by the atmospheric turbulence on the wavefronts coming from the stars. Without AO techniques the atmospheric turbulence limits the angular resolution of images to that obtained with a $10$ cm telescope. To correct the turbulence we need to know it. For this reason the characterization of the optical turbulence ($\CN2$ profiles and integrated values) is a crucial element to guarantee successful ground-based observations and feasibility of ELTs. Today astronomers need to know how much turbulence is developed above an observatory and how the turbulence is distributed in the troposphere ($\sim$ $20$ km). This can support, in a suitable way, the selection and scheduling of scientific programs to be realized with different kind of instruments placed at the focus of telescopes (flexible-scheduling), the selection of the best sites for new telescopes and also the employment of the most recent AO techniques such as the Multi-Conjugated Adaptive Optics (MCAO), the Laser Guide Star (LGS) and the tomographic technique.  In spite of the fact that all projects of new ELTs plan to have a flexible-scheduling system to optimize the management of telescopes, no telescope, at present, employs a real flexible-scheduling system based on the forecast of the optical turbulence. At the same time, an efficient search for and selection of the best astronomical sites in the world is difficult to be realized because measurements are not homogeneous, the site testing campaigns are quite expensive and the required time to characterize a site with measurements can extend to decades.\newline

A few instruments, able to measure one or a few of the astroclimatic parameters characterizing the wave-front perturbations induced by the atmospheric turbulence, exist at present. All these instruments provide {\bf local} measurements that can be taken {\it in situ}, as is the case of radiosoundings and instrumented masts, or along a single line of sight, as is the case of optical instruments such as the Generalized Scidar (GS)\cite{VerninAzouit1983,Avila1997}. The numerical technique, more precisely the simulations of the $\CN2$ provided by atmospherical non-hydrostatic meso-scale models (Meso-Nh)\cite{Masciadri1999a,Masciadri1999b,Masciadri2001a,Masciadri2004} offers, with respect to measurements, a few fundamental advantages: {\bf (1)} the possibility to provide 3D $\CN2$ maps above a region having a surface of a few tens of kilometres around a telescope;  {\bf (2)} the possibility to simulate the $\CN2$ {\it  'where'} and {\it 'when'} it is required without the necessity of long and expensive site testing campaigns done with several instruments and this is particularly interesting for searches of new sites;  {\bf (3)} the possibility to forecast $\CN2$ profiles, (a real {\it 'chimera'} for all astronomers), fundamental information for the flexible scheduling of the scientific programs. Astronomical observations related to the most challenging scientific programs frequently can be realized only with excellent turbulence conditions. Our ability in forecasting these particular conditions is thus determinant to progress in astrophysics knowledge. To give a clear idea of the scientific impact of the employment of the flexible scheduling obtained with the forecast of the optical turbulence (OT) I recall that, in 1984, Roddier and Lena\cite{RoddierLena1984}showed that, passing by a $r_{0}$ of $10$ cm to an $r_{0}$ of $20$ cm (equivalent of an improvement of the seeing from $1$ to $0.5$ arcsec) we gain a factor $2$ in the magnitude limit of observed scientific targets. It is possible to show\cite{Veron1996} that, passing by a magnitude limit of $12$ to $14$ in the visible, the number of Active Galactic Nuclei (AGN) detectable from the ground and having a declination between -$70$$^{\circ}$ and +$10$$^{\circ}$, increases from $5$ to $100$. The impact of the scientific feedback obtained with these studies is therefore particularly important. \newline

\begin{figure*}
\centering
\includegraphics[width=7cm]{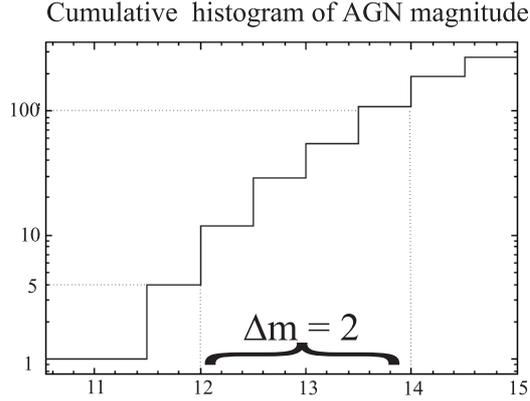}
\caption{Cumulative histogram of AGN magnitudes $V$ ($V$ $\leq$ 15) for a declination in 
the [$-70$, $+10$] degrees range (V\'eron-Cetty \& V\'eron, 1996)}
\label{win_sum_wind}
\end{figure*}

The reconstruction of the optical turbulence done with atmospherical models is extremely important not only for the {\it forecasting} but also in different other contexts. I remind here a few of these contexts trying to put in evidence the potentiality of these studies in future applications to the High Angular Resolution.  \newline

During the past few years it was proven that the limitations of the AO techniques due to the turbulence change as the AO techniques evolve and proceed\footnote{Ragazzoni et al.,\cite{Ragazzoni2000} for example, proved that the wavefront perturbations at an altitude $h$ can be described by a modal tomographic matrix M that depends on the altitude $h$ and the relative strength of the perturbed layers in the whole atmosphere $M$$=$$M($\CN2$)$. The typical lifetime $\tau_{M}$ of the matrix M should be therefore estimated to optimize this technique.}. For this reason, it would be quite useful to develop a flexible tool that can characterize the turbulence in an exhaustive way and that can easily be modified in order to satisfy new requirements of the AO techniques and the site testing without the necessity to conceive and build new dedicated instruments. The numerical technique is the only tool having such flexibility. In application to a GLAO system, for example, it might be relatively easy to increase the vertical resolution of a model in proximity of the ground to better reconstruct the turbulence distribution and resolve thin layers (hundreds or tens of meters) in the first kilometer. A GLAO system, indeed, is a wide field AO system and is therefore characterized by a short depth of field. For this reason to optimize a GLAO system we necessarily need, at the same time, to characterize the turbulence with a vertical resolution of the order of few tens of meters and to know this vertical distribution at different lines of sight. Due to the presence of orographic effects the approximation of uniform horizontal distribution of the $\CN2$, normally assumed for most of applications to the AO, might be no more acceptable and might badly describe the reality if we look at the turbulence with a high vertical resolution in the surface and boundary layer. 3D $\CN2$ maps might be therefore fundamental to optimize a GLAO system.\newline 

In the last years astronomers discovered that the climatology of the optical turbulence is a quite critical information that necessarily should be taken into account in the selection of the best astronomical sites. It is well known the case of Cerro Paranal, selected in the 1990 as the best site for the VLT. Drastic changes in the climatology of meteorologic parameters above the Cerro Paranal Chilean region produced a deterioration of the quality of this site characterized, at present, by a worse median seeing than during the nineties. The climatology of the optical turbulence (as the climatology of whatever meteorological parameter) has to be necessarily done with atmospherical models particularly if we want to access it on whatever geographic coordinates above the earth. The atmospheric models, differently from measurements, permits us to extend our knowledge to the past that is to access something happened in the past. The possibility to reconstruct the turbulence climatology vertical distribution in the past can permit us to better evaluate and understand the turbulence characteristics above whatever site and to extrapolate its future trends. At present, the non-hydrostatics meso-scale models are the best tool that might produce a climatology of the optical turbulence extended to decades. Only these models, indeed, can resolve the typical high spatial and temporal fluctuation scales characterizing the optical turbulence. \newline

The financial resources invested in ground-based astronomy are huge and only an efficient management of these structures and instruments can get ground-based astronomy competitive versus the spatial one. The success of ELTs will pass necessarily by the implementation of a reliable and efficient flexible-scheduling system and the optical turbulence forecast made by atmospherical models will be necessarily a crucial element in the future operation strategy of new generation ground-based telescopes. 

The numerical technique is, at present, the only one that can provide 3D $\CN2$ maps and the forecast of the optical turbulence. It is the only technique able to potentially characterize a huge number of sites in the world, permitting an efficient selection done in homogeneous way. However, the ability in {\bf forecasting} the $\CN2$ maps has not yet been proved; this justifies the need for further studies in this field. \newline

A project, FOROT, has been recently funded by the European Community (Marie Curie Excellence - FP6 Program) to boost studies of the optical turbulence forecast. In Section \ref{art} I will summarize the status of art of our knowledge and the main results already obtained in this field. In Section \ref{scient_tec} I will summarize the scientific and technological objectives of FOROT. In Section \ref{syner} I will give a general panorama of international synergies that we joint to carry out this project. In Section \ref{conc} I will present my conclusions.

\section{Status of Art}
\label{art}

Chris Coulman\cite{Coulman1986}(1986) was the first to exhort collaborations between astronomers and meteorologists to solve the problem of the forecast of the optical turbulence. It was only in $1995$ that this invitation was accepted\cite{Bougeault1995}.
 The authors\cite{Bougeault1995} proposed to use a hydrostatic model (PERIDOT) 
developed in the Centre 
National des Rech\`erches M\'et\'eorologiques (CNRM/M\'eteo France - Toulouse, Fr) 
linked 
with an orographic model having a horizontal resolution of $3$ and $10$ km 
to resolve the orographic waves produced by the friction of the atmospheric flow over the ground. 
Good qualitative results were obtained but the turbulence was, in general, 
underestimated because of a relatively low horizontal resolution. 
In the same year a different approach has been proposed\cite{DeYoung1995} 
using an orographic model and atmospheric model 
having a very high resolution (about $10$ m) extended over a surface of 
a few hundreds of meters. They tested their model over the Mauna Kea and Cerro 
Pach\'on sites. The limit of this work was an idealized initialization of the model 
and a simplified turbulent equation that did not consider the buoyancy 
forces which are the principal cause of the production of gravity waves.\newline

In $1999$ it was proposed\cite{Masciadri1999a,Masciadri1999b} to use the non-hydrostatic 
model Meso-Nh linked with an orographic model having an horizontal resolution higher than $1$ km. For the first time the validation of an atmospherical model was done comparing directly the measured and simulated $\CN2$ profiles. Since these first results I could obtained new results and achieved better performances of the Meso-Nh model. Which is the status of art at present ?\newline

\begin{figure}
\centering
\includegraphics[width=9cm]{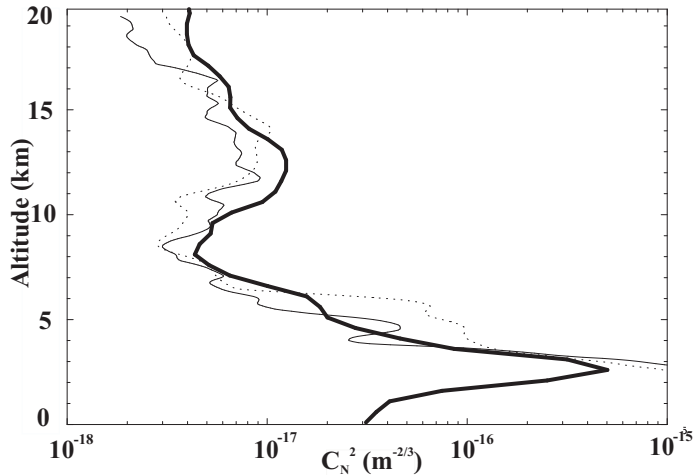}
\caption{Mean vertical $\CN2$ profiles measured and simulated over the 
whole SPM2000 campaign. Bold line: GS. Thin line: radiosoundings. Dotted line: Meso-Nh model. The relative total seeing obtained from the integration of the $\CN2$ profiles is: $0.79$ arcsec (GS), $1.07$ arcsec (balloons), $0.93$ arcsec (Meso-Nh model). 
\label{cn2_mnh_sci_bal}} 
\end{figure}

The main scientific results obtained so far and starting point for the project FOROT are:
\begin{itemize}

\item We could prove that the parametrization of the $\CN2$ is possible, that means that we are able to produce reliable $\CN2$. 
This result was not evident a priori due to the fact that the optical turbulence has typical spatial and temporal scales of fluctuation smaller than those of the standard meteorological parameters and it can not be explicitly resolved by the model over the whole troposphere.
\item We validated the model above two among the best astronomical sites in the world (Cerro Paranal\cite{Masciadri1999a,Masciadri1999b} and Roque de los Muchachos\cite{Masciadri2001a}). We proved that measured and simulated $\CN2$ profiles well match from a {\it qualitative} (comparable shape) and {\it quantitative} (comparable integral over the $25$ km) point of view. At the same time we put in evidence some limitation and possible improvements to be implemented in the model. 
\item A new method\cite{MasciadriJabouille2001} to calibrate the Meso-Nh model has been proposed to eliminate some potential systematic errors and improve the reliability of the model.
\item More recently the Meso-Nh model has been statistically validated\cite{Masciadri2004} comparing simulated and measured $\CN2$ profiles obtained during the site testing campaign SPM2000 above San Pedro M\'artir (Mexico)\cite{Avila2004}. Figure \ref{cn2_mnh_sci_bal} shows the $\CN2$ vertical profiles simulated by the Meso-Nh model and measured by a Generalized Scidar and balloons. These are mean values obtained over $10$ nights. The main and fundamental result obtained in this study is that we showed that the dispersion between simulations and measurements ($\Delta \varepsilon_{GS,model}$) is comparable to the dispersion between measurements provided by different instruments ($\Delta \varepsilon_{GS,balloons}$): 
\begin{equation}
\Delta \varepsilon_{GS,model} = \Delta \varepsilon_{GS,balloons} \sim 30 \%
\label{eq1}
\end{equation}
This result showed that the numerical technique can be considered, at present, a reliable technique to estimate $\CN2$ profiles and, after a calibration, it can be used in an autonomous way to simulate $\CN2$ profiles. I would like to underline a point that is fundamental from the point of view of perspectives of the basic research in this field. The strength of the result summarized in Eq.\ref{eq1} is that, in spite of the fact that performances of this model can be improved in the future, at present we can state that the reconstruction of the optical turbulence made with an atmospherical model such as Meso-Nh is done with an accuracy (dispersion from the true) that is not worse than what obtained by measurements. 

For the not-experienced reader it is worth to underline that we are talking here about statistical comparison i.e. median and mean values on one or a set of nights. No comparison between measurements and simulations is possible, at least at present, on time scale of minutes. 

\item We completed the first seasonal variation study\cite{MasciadriEgner2005,MasciadriEgner2005bis} of the vertical distribution of the optical turbulence ($\CN2$ profiles) and all integrated astroclimatic parameters ($\varepsilon$, $\theta_{0}$, $\tau_{0}$, $\theta_{M}$, $\LO$) above the San Pedro M\'artir site (Baja California). This study has been particularly important because permitted us to show the real potentialities of the simulations. $80$ $\CN2$ and wind speed vertical profiles uniformly distributed in one year permitted us to calculate all the derived astroclimatic parameters and even the optimal altitude at which to conjugate the DMs of a MCAO system having 1, 2 and 3 DMs. The numerical technique permitted us to put in evidence the gain obtained in changing the optimal conjugation heights of the deformable mirrors of a MCAO system in winter and summer time.

\item We could prove that meso-scale models can discriminate and identify sites having particularly weak wind speed near the ground while GCMs can not\cite{Masciadri2003}. Moreover, the relative error $\Delta r$ between the measured and reconstructed wind speed near the ground obtained by meso-scale models is 27-47 $\%$ smaller than what obtained by the General Circulation Models such as those of the ECMWF. These last are weakly reliable in proximity of the surface due to the fact that orographic effects are badly represented by such a models. 
\end{itemize}

\section{Scientific and Technological Goals}
\label{scient_tec}

\begin{figure}
\centering
\includegraphics[width=9cm]{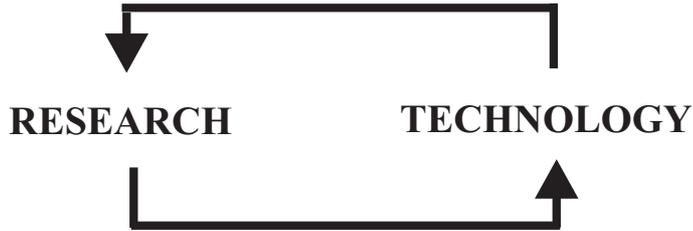}
\caption{Schematic diagram showing the nature of mutual interactions between research and technology (see text).
\label{res}} 
\end{figure}

This discipline necessarily needs to pass now to a new dimension to completely exploit its potentialities. This leap ahead has to be done in two directions: research and technology. The FOROT Project moves from these motivations.\newline\newline
\noindent
{\bf Research} The project FOROT (funded by the FP6 Program - Marie Curie Excellence) is conceived to reach two main scientific goals: {\bf (1)} to prove/test the ability of the non-hydrostatic meso-scale models in forecasting the optical turbulence and {\bf (2)} to prove/test the ability of these models in discriminating the quality (from the turbulence point of view) of astronomical sites. I remind that we are talking of differences of the order of decimal fractions of arcseconds.
To concretize our studies and better focalize our objectives we decided to apply this technique in two different regions of the world.
\begin{itemize}
\item {\bf Mt. Graham (Arizona)}, site of the Large Binocular Telescope.\newline 
{\it Main Scientific goal:} to test the model ability in forecasting the optical turbulence.
\item {\bf Antarctic Plateau (Dome C, South Pole and Dome A)}.\newline
{\it Main Scientific goal:} to test the model ability in discriminating sites above the Antarctic Plateau.
These three pre-selected sites are well known by the astronomical community to be among the best on this continent for astrophysical science. At South Pole a base exists since several years. At Dome C, an Italo-French base (Concordia) is being built and a set of projects of small-medium size telescopes and interferometers is planned by several international consortia. Dome A has not yet been explored and monitored so far but it might be even better than Dome C due to a higher altitude of the site. The main idea is to validate and calibrate the model above Dome C and South Pole using the measurements done so far above these two sites. After the calibration, the model should be able to simulate the turbulence above Dome A with a high level of reliability.
\end{itemize}
\noindent
{\bf Technology} The numerical technique applied to astronomy needs, to be effective, to pass to a technological development aiming to get operational a nightly forecasting of the optical turbulence above an astronomical site. In this project we plan to install the Meso-Nh model on an autonomous machine (instead of supercomputers of meteorologic centers) to be able to run in a systematic way the model above the astronomical sites. This will allow us to do a great leap in the direction of an implementation of the real flexible-scheduling system but still remaining in a research dimension. Astronomers need to appropriate of a tool (atmospherical model) usually employed by meteorologists. A careful work in the development of interfaces between the autonomous machine and the data archives of the European Center for Medium Weather Forecasts (ECMWF) has to be done to permit us to initialize the Meso-Nh model with analyses and forecasts produced by the General Circulation Models from ECMWF. As Fig.\ref{res} shows, a leap in the technology field can trigger further studies and get more challenging some scientific goals. An efficient closed loop can be installed between research and technology promoting progress in both directions.

\section{International Synergies}
\label{syner}

Measurements are an important element for
studies of the atmospheric turbulence made with 
atmospheric models. Measurements are crucial to validate this kind of models
and to supply strength to the numerical methodology. In our project (FOROT) we tried to catalyze international 
synergies in order to collect existent measurements and to assure a future supply of critical measurements.

Above Mt. Graham, since a couple of year, we have started to run a Generalized Scidar (GS/LBT) at the focus of the VATT ($1.75$ m) placed some tens meters far away from the LBT. $\CN2$ profiles related to $16$ nights\cite{Egner2006} have been collected so far. Further GS/LBT runs are planned in the future months/years and will be taken care by the FOROT Group. This activity has been supported by the VATT staff and the LBT Consortium (Steward Observatory-Arizona, INAF/Arcetri-Italy, MPIA-Germany). Two anemometers, placed on the roof of the LBT and beside the dome of the VATT, have permitted us to validate the wind speed measured by the GS/LBT near the ground. Further instrumentation, mainly addressed to turbulence measurements, is planned in the near future to be tested/installed above Mt. Graham.

Above the Antarctic Plateau we will be able to use a set of optical turbulence measurements of different nature ($\CN2$ vertical profiles measured with radiosoundings and optical instruments, integrated values such as seeing $\varepsilon$ and isoplanatic angle $\theta_{0}$) that during the last decade have been done by a few teams who are involved in site testing studies above the Antarctic Plateau. Among these the LUAN Team (Nice, France) and the Team of the New South Wales University (Sydney, Australia). The French group will provide us $\CN2$ vertical profiles taken by a Single Star Scidar (SSS) and radiosoundings, measurements of seeing ($\varepsilon$) and isoplanatic angle ($\theta_{0}$) provided by DIMMs placed at different altitudes from the ground\cite{Aristidi2005,Agabi2006}. All these measurements have been taken above Dome C during the summer and winter time. 
The Australian group will provide us meteorological parameters and $\CN2$ profiles taken above South Pole some years ago\cite{Marks96,Marks99}. Above Dome C it is possible to register, in this moment, several activities related to the site testing. We hope/expect therefore to include in the set of measurements of reference further ones such as those obtained by a lunar SHABAR\cite{Moore2006} (INAF Arcetri, Italy/CELT, USA).\newline 

In the context of the measurements/simulations (from atmospherical models) comparison it is worth to mention that the level of accuracy reached so far in estimating and quantifying the atmospheric optical turbulence employing different instruments is of the order of $20$-$30$ $\%$. As already said in Section \ref{art}, indeed, a discrepancy of $\sim$$30$$\%$ has been found between the seeing measured by balloons and a GS\cite{Masciadri2004} above San Pedro M\'artir. A similar value ($\sim$$30$$\%$) has been found above Cerro Tololo\cite{Vernin1998} between balloons and a GS. Balloons provide $\CN2$ vertical profiles with an intrinsic accuracy of the order of $20$-$30$$\%$ as the same authors declare\cite{Vernin2005}. If we compare optical instruments (vertical profilers) such as the GS and the MASS we find a dispersion of the seeing ($\varepsilon$) of $\sim$$20$$\%$ in the range $8$-$16$ km\cite{Toko2005}. However, in the first $4$ km, the typical dispersion of the seeing ($\varepsilon$) can reach values of the order of $\sim$$50$-$100$$\%$\cite{Toko2005}. Of course, this does not want to be a criticism of measurements but only a definition of the accuracy with which we can estimate, at present, the turbulence. This is a crucial element if one intends to reconstruct the turbulence with an atmospheric model and define a score of success.\newline  
If we look at instruments measuring integrated values of the turbulence, things are not different. Recently\cite{Agabi2006}, values of the isoplanatic angle ($\theta_{0}$) have been measured by a DIMM ($4.7$ arcsec) and balloons ($2.7$ arcsec) above Dome C in the Antarctic Plateau in the same period during the winter time. The discrepancy of these measurements is of the order of $\sim$ $50$$\%$. This dispersion is quite worried because, due to the fact that $\theta_{0}$ is weakly dependent from the turbulence in the ground layer, $\theta_{0}$ should be mostly affected by the turbulence in the free atmosphere. The height at which the DIMM is placed and the spurious values provided by the balloons in the first tens meters should therefore weakly affected the $\theta_{0}$.  There are no reasons therefore that these two results are so different even taking into account the factors just described (height of the DIMM and spurious values provided by the balloons in the first tens meters). Such a large discrepancy suggests the necessity of further measurements to dissipate doubts, verify that no bias are included in one of two or both instruments and definitely estimate $\theta_{0}$ in winter time.

Several factors contribute to this not negligible dispersion between measurements but, as we explained with the previous examples, is certainly not only due to a statistical uncertainty. Among the causes: systematic errors, intrinsic complexity of the phenomenon itself, techniques of measurements strongly different that sometime get hard any comparison between measurements provided by different instruments. As a consequence also comparisons between measurements and simulations has to be done quite carefully. It is fundamental to use as many as possible measurements taken contemporary and provided by different instruments in order to define a sort of {\it accuracy of reference} that can be compared to the accuracy of simulations. It is evident that this is necessary to define a {\it score of success} of simulations. We exhort therefore, for the present and the future, a quite tight collaboration between the FOROT group and the other teams involved so far in site testing activity in trying to jointly planning campaigns of measurements taking into account the necessity to define an {\it accuracy of reference}. This implies the necessity of more site testing campaigns aiming to quantify the turbulence but also to cross-check instruments to understand {\it where} there are still problems/uncertainties and how to solve them. This should provide a useful information not only for studies related to the turbulence forecast but to the whole ground-based astronomy.

\section{Conclusion}
\label{conc}

In this paper the main scientific and technological goals of the project FOROT are presented. FOROT finds its main motivations on fundamental challenges for the success of the future ground-based astronomy. I described how the reconstruction of the optical turbulence above astronomical sites done with atmospherical models at meso-scale can be useful to answer to major open questions in the field of the characterization of the optical turbulence and represents the natural answer to the need of the turbulence forecast and turbulence climatology. We will not have a real flexible-scheduling system without turbulence forecast.\newline
The road for this final goal pass necessarily by a further development of this discipline at a basic research level and we need to catalyze international synergies to maximize the success of this project and put the bases for a future of this discipline. FOROT intends to be a first step on this road-map.

\acknowledgments        

This work is funded by a Marie Curie Excellence Grant (FOROT) 
MEXT-CT-2005-023878 - FP6 Program.\newline
I thanks all the collaborators who participated, at different levels, to these researches in the past:
Jean Vernin, Philippe Bougeault, Patrick Jabouille, Remy Avila, Leonardo Sanchez, Tania Garfias, Sebastian Egner and Kerstin Geissler. \newline 
I also would like
 to thanks a few scientists who supported these unusual researches: Irene Cruz Gonzalez and Piero Salinari.


\end{document}